\documentclass[american,aps,prl,reprint,superscriptaddress,twocolumn]{revtex4-1}
\usepackage[unicode=true,pdfusetitle, bookmarks=true,bookmarksnumbered=false,bookmarksopen=false, breaklinks=false,pdfborder={0 0 0},backref=false,colorlinks=false] {hyperref}
\hypersetup{ colorlinks,linkcolor=myurlcolor,citecolor=myurlcolor,urlcolor=myurlcolor}
\usepackage{braket,colortbl,cleveref,amsthm,amsmath,amssymb,txfonts}
\definecolor{myurlcolor}{rgb}{0,0,0.7}
\usepackage{graphics,graphicx}
\usepackage{color}

\theoremstyle{plain}

\def\bea{\begin{eqnarray}}
\def\eea{\end{eqnarray}}
\def\ba{\begin{array}}
\def\ea{\end{array}}

\def\ket{\rangle}
\def\bra{\langle}
\def\beq{\begin{equation}}
\def\eeq{\end{equation}}

\begin{document}

\title{ Convex resource theory of non-Markovianity}

\author{Samyadeb Bhattacharya} 
\email{sbh.phys@gmail.com}
\author{Bihalan Bhattacharya}
\email{bihalan@gmail.com}
\author{A. S. Majumdar}
\affiliation{S. N. Bose National Centre for Basic Sciences, Block JD, Sector III, Salt Lake, Kolkata 700 098, India}

\begin{abstract}
\noindent We establish a convex resource theory of non-Markovianity under the constraint of small time intervals within the temporal evolution. We construct the free operations, free states and a generalized bona-fide measure of non-Markovianity. The framework satisfies the basic properties of a consistent resource theory. The proposed resource quantifier is lower bounded by the optimization free Rivas-Huelga-Plenio (RHP) measure of non-Markovianity. We further define the robustness of non-Markovianity and show that  it can directly be expressed as a function of the RHP measure of non-Markovianity. This enables a physical interpretation of the RHP measure.

\end{abstract}

\maketitle

\textit{Introduction:} Control and manipulation of characteristic quantum traits of any physical system are more often than not hindered by decoherence resulting from unavoidable coupling with noisy environments. Theory suggests that as a result of decoherence, the system monotonically relaxes to thermal equilibrium, or generally, to a non-equilibrium steady state \citep{alicki,lindblad,gorini,breuer}. The one-way information flow characterized by the monotonic relaxation towards the stationary states is a direct consequence of the Born-Markov approximation \citep{breuer}, which is valid for very large stationary environments, leading to Complete Positive (CP)-divisibility of the dynamics \citep{rivas1,breuerN,alonso}. However, beyond Born-Markov limit, the CP-divisibility breaks down \citep{rhp1}, triggering non-Markovian (NM) backflow of information \citep{blp1,proman1,chanda,proman2,proman3,proman4,proman5,proman8,proman9,bellomo,arend,ban1,ban2,samya1,samya2,proman10}. 

Recently, it has been shown that NM information backflow acts as a resource in various quantum mechanical tasks. For example, it has been shown that NM allows perfect teleportation with mixed states \citep{task1}, efficient entanglement distribution \citep{task2}, improvement of capacity for long quantum channels \citep{task3} and efficient work extraction from an Otto cycle \citep{task4}. For all of these mentioned cases, the accomplishment of the tasks has been done by harnessing information backflow, which can be understood as resource inter-conversion. NM can thus be inter-converted via information backflow, into other resources like entanglement, coherent information, extractable work etc. It can also be exploited for efficient quantum control \citep{task5}. Thus it is evident, that we are in need of a proper resource theoretic framework of NM in quantum theory. Furthermore, we know that information theories can be viewed as examples of resource inter-conversion \citep{M2}. Since NM can be converted into other resources, the necessity of a resource theory of non-Markovianity (RTNM) is strongly established. 

The  construction of resource theories in connection with various quantum phenomena such as entanglement \citep{Re1,Re2}, coherence \citep{Re4}, reference frame and asymmetry \citep{Re3}, nonlocality \citep{Re7}, non-gaussianity \citep{Re8}, and thermodynamics \citep{Re5,Re6}, has flourished in recent years. In this letter we construct a RTNM of similar structure, by developing  its fundamental components. Previously, there has been an attempt to construct a classical RTNM \citep{rtnm1} based on a tripartite scenario. The phenomenon of NM is of course not restricted within quantum theory only \citep{example1}. However, information backflow \citep{blp1} is an explicitly quantum characteristic. Since our RTNM encapsulates information backflow caused by indivisibility of quantum channels as the central feature, it can be understood as a quantum framework of RT, valid for arbitrary finite dimensional single or \textit{any}-partite system, satisfying all the basic ingredients \citep{rt1}. Notably, there are arguments on whether or not the divisible operations exhaust all the Markovian operations \citep{modi}. However, even if there exists CP-divisible NM operations, they will not generate information backflow. Hence it is legitimate to consider only indivisible operations as resourceful operations.

Note that, our construction of RTNM is restricted to the operations having Lindblad type generator $\dot{\rho}(t)=\mathcal{L}(\rho(t))=\sum_{\alpha=1}^{n\leq d_S^2}\Gamma_{\alpha}(t)\left(L_{\alpha}\rho(t)L_{\alpha}^{\dagger}-\frac{1}{2}L_{\alpha}^{\dagger}L_{\alpha}\rho(t)-\frac{1}{2}\rho(t)L_{\alpha}^{\dagger}L_{\alpha}\right)$, where $\Gamma_{\alpha}(t)$s are the Lindblad coefficients, $A_{\alpha}$s are the Lindblad operators and $d_s$ is the dimension of the system. For divisible evolutions $\Gamma_{\alpha}(t)\geq 0,~ \forall \alpha, t$. 

\textit{Resource theory of non-Markovianity:}
There are many types of physical phenomena in quantum mechanics, defined directly at the level of quantum states by imposing constraints over the physical operations. To construct a RT concerning one of such phenomena, we need to identify the states containing the signature of such resources; or conversely, we need to find the states not containing such resources, defined as the ``free states". In case of NM, since it is a property of quantum processes, constructing free and resourceful states is not unequivocal. The second important component of any quantum RT is the set of constrained quantum operations under which the resource cannot increase. These operations are called the ``free operations". 
Thirdly, we need a resource quantifier, which is contractive under free operations. Additionally, it is technically convenient, if the RT structure is convex. Quantum RTs such as entanglement, coherence, asymmetry and athermality possess such structure. In the following we construct these basic components for the RTNM and prove that it satisfies all the requirements of a convex RT under a particular constraint. 

\textbf{Free operations:} We define, the divisible operations as the resource non-increasing operations. They are expressed as 
$\Lambda^{\mathcal{M}}(t_2,t_1)\equiv \exp\left(\int_{t_1}^{t_2}\mathcal{L}_tdt\right),$
where
$
\mathcal{L}_t(\cdot)=\sum_i \Gamma_i(t)\left(L_i(\cdot)L_i^{\dagger}-\{L_i^{\dagger}L_i,(\cdot)\}\right)~with~\Gamma_i(t)\geq 0, \forall i,t.
$

\textbf{Free states:} As discussed earlier, since NM is solely the property of underlying operation, there is no concept of ``states" for it. But to utilize the geometry of state space underlying the structure of RT, it is necessary to construct the set of free states corresponding to a particular resource. We bridge this gap by considering the Choi states \citep{choi,jamil} corresponding to the free operations as free states. They are defined as 
$\mathcal{C}^{\mathcal{M}}(t+\epsilon,t)=\mathbb{I}\otimes\Lambda^{\mathcal{M}}(t+\epsilon,t)(|\psi\ket\bra\psi|)~~with~~\epsilon>0,$
where $|\psi\ket$ is the maximally entangled state of $d\times d$ dimension, for a $d$ dimensional system.
Recall the RHP measure of NM \citep{rhp1} given as
$
g^{\mathcal{N}}(t)=\lim_{\epsilon\rightarrow 0^+}\dfrac{\parallel \mathcal{C}^{\mathcal{N}}(t+\epsilon,t)\parallel_1-1}{\epsilon},
$
where $\parallel\cdot\parallel_1$ is the trace norm  and $\mathcal{C}^{\mathcal{N}}(t+\epsilon,t)=\mathbb{I}\otimes\Lambda^{\mathcal{N}}(t+\epsilon,t)(|\psi\ket\bra\psi|)$ is the Choi state corresponding to any operation $\Lambda_{\mathcal{N}}$. For a divisible evolution, $g^{\mathcal{N}}(t)$ is always zero since $\mathcal{C}^{\mathcal{M}}(t+\epsilon,t)$ is valid state and $\parallel \mathcal{C}^{\mathcal{M}}(t+\epsilon,t)\parallel_1=1$. Thus we define the set of free states as 
\[\mathbb{F}=\left\{~\mathcal{C}^{\mathcal{M}}(t+\epsilon,t)~~|~~\parallel \mathcal{C}^{\mathcal{M}}(t+\epsilon,t)\parallel_1=1~\forall~\epsilon,t~\right\}.\]
We prove the following propositions stating particularly essential properties. 

\noindent\textbf{Proposition 1A:} \textit{Resourceful states cannot be generated under tensor product, partial trace and permutations of spatially separated subsystems.}

\proof  Let us consider two arbitrary Choi states $\rho^\mathcal{M}$, $\sigma^\mathcal{M}$ $\in$ $\mathbb{F}$. Therefore $\parallel \rho^\mathcal{M} \parallel_1=\parallel \sigma^\mathcal{M} \parallel_1=1$. From the properties of trace norm \citep{trace1} we have, 
$\parallel \rho^\mathcal{M} \otimes \sigma^\mathcal{M}\parallel_1 = \parallel \rho^\mathcal{M} \parallel_1 \parallel \sigma^\mathcal{M} \parallel_1=1$. We know that for a NM resourceful Choi state, the trace norm must be strictly greater than 1 in some intermediate region.
Therefore $\rho^\mathcal{M} \otimes \sigma^\mathcal{M}$ is not a resourceful state. 

Similarly we can prove $\sigma^{\mathcal{M}}\otimes\rho^{\mathcal{M}}$ is also not resourceful. \\
To show partial trace cannot generate resource, let $\rho^{\mathcal{M}}_{AB}$ $\in$ $\mathbb{F}$ be an arbitrary bipartite Choi state. If we take partial trace with respect to subsystem B, the reduced subsystem becomes
$\rho^{\mathcal{M}}_{A}= Tr_{B}[\rho^{\mathcal{M}}_{AB}]$.
Now, $\parallel \rho^{\mathcal{M}}_{A} \parallel_1$ = Tr[$\sqrt{\rho^{\mathcal{M}}_{A} \rho^{{\dagger}\mathcal{M}}_{A}}$]= Tr[$\rho^{\mathcal{M}}_{A}$], as $\rho^{\mathcal{M}}_{A}$ is a positive operator. Hence we have
\[\rho^{\mathcal{M}}_{A} = \sum_{j=1}^{d_B} (\mathbb{I}_A\otimes \langle b_i \vert) \rho^{\mathcal{M}}_{AB} (\mathbb{I}_A \otimes \vert b_i \rangle),\]
where $d_B$ is the dimension of subsystem B having orthonormal basis $\{|b_i\ket\}$. Therefore,
\[
\begin{array}{ll}
Tr[\rho^{\mathcal{M}}_{A}]=Tr[\rho^{\mathcal{M}}_{AB}(\mathbb{I}_A \otimes \sum_{j=1}^{d_B} \vert b_i \rangle\langle b_i \vert)]
=Tr[\rho^{\mathcal{M}}_{AB} (\mathbb{I}_A \otimes \mathbb{I}_B)]\\
= Tr[\rho^{\mathcal{M}}_{AB}] = 1.
\end{array}
\]
So we prove that $\rho^{\mathcal{M}}_{A}$ is not a resourceful state.    \qed

\noindent\textbf{Proposition 1B:} \textit{The set of free states $\mathbb{F}$ is a compact set.}

\proof  A subset in an Euclidean space is compact if and only if it is bounded and closed (contains all its limit points). Boundedness of $\mathbb{F}$ in trace norm is clear from its definition and closedness of $\mathbb{F}$ follows from the continuity of trace norm. Hence $\mathbb{F}$ is compact. \qed

\noindent\textbf{Proposition 1C:} \textit{Free operations cannot generate resourceful states.}

\proof Here we need to prove that free operation $\Lambda^{\mathcal{M}}$ cannot generate resourceful state. i.e. if $\sigma(t+\epsilon,t)=\mathbb{I}\otimes\Lambda^{\mathcal{M}}(t+\epsilon,t)\mathcal{C}^{\mathcal{M}}(t)$, then $\sigma(t+\epsilon,t)\in \mathbb{F}$. i.e. $\parallel\sigma(t+\epsilon,t)\parallel_1=1,~\forall \epsilon, t$.\\
Now $\Lambda^{\mathcal{M}}(t+\epsilon,t)$ is a divisible CPTP map and the free state $\mathcal{C}^{\mathcal{M}}(t)$ is a valid density matrix. Therefore the divisibility of $\Lambda^{\mathcal{M}}(t+\epsilon,t)$ implies that it will take valid density matrix $\mathcal{C}^{\mathcal{M}}(t)$ to a valid density matrix $\sigma(t+\epsilon, t)$ $\forall t,\epsilon$. Therefore, from the property of density matrices, we have $\parallel\sigma(t+\epsilon,t)\parallel_1=1~\forall t,\epsilon$. \qed

\noindent The implications of these propositions are  important. Free states are free in all possible finite dimensions. One cannot generate resourceful states without any cost. Therefore it is natural that tensor product of two free states and reduced state of a free state cannot be resourceful \citep{rt1} and free operations cannot generate resource.

\noindent\textbf{Measure of non-Markovianity:} A proper measure of any resource can be constructed by the minimum contractive distance between a resourceful state and the set of free states.  Here we have to consider the distance between a Markovian and a non-Markovian operation. Choi-Jamilkowski isomorphism \citep{choi,jamil} reduces this problem to finding distance between corresponding Choi states. A measure of NM can be defined as $M(t+\epsilon,t)=\inf_{\mathcal{C}^{\mathcal{M}}(t+\epsilon,t)}D(\mathcal{C}^{\mathcal{N}}(t+\epsilon,t)|\mathcal{C}^{\mathcal{M}}(t+\epsilon,t))$, where $D(\cdot|\cdot)$ is any metric contractive under CPTP maps. $M(t+\epsilon,t)$ can only be non-zero positive quantity in the time span $\epsilon$, where CP breaks down. The optimization is done over the free states ($\mathcal{C}^{\mathcal{M}}(t+\epsilon,t)$). This is extremely 
 difficult to calculate because the free operations $\Lambda^{\mathcal{M}}$ do not form a convex set \citep{wolf1,wolf2}. We overcome this difficulty by virtue of the following proposition.

\noindent\textbf{Proposition 2:} \textit{In the limit of $\epsilon\rightarrow 0$, the Choi states $\mathcal{C}^{\mathcal{M}}(t+\epsilon,t)$ form a convex set.}

\proof To prove the convexity of a set, it is enough to show that the convex combination of two elements of the set also belongs to the same set. Let us consider two Markovian operations 
\[\Lambda^{\mathcal{M}}_{(1)}(t+\epsilon,t)= \exp\left(\int_{t}^{t+\epsilon}\mathcal{L}_t^{(1)}dt\right),
\Lambda^{\mathcal{M}}_{(2)}(t+\epsilon,t)\equiv \exp\left(\int_{t}^{t+\epsilon}\mathcal{L}_t^{(2)}dt\right)\] 
Here $\mathcal{L}_t^{(1)}$ and $\mathcal{L}_t^{(2)}$ are Lindblad type generators with positive coefficients. In the limit of $\epsilon\rightarrow 0$, we can expand the exponentials and neglect the 2nd and higher order terms. Therefore, we have 
\[\Lambda^{\mathcal{M}}_{(1)}(t+\epsilon,t)=\mathbb{I}+\epsilon\mathcal{L}_t^{(1)},
\Lambda^{\mathcal{M}}_{(2)}(t+\epsilon,t)=\mathbb{I}+\epsilon\mathcal{L}_t^{(2)}.\]
We define another map
\[
\begin{array}{ll}
\Lambda^{\mathcal{M}}(t+\epsilon,t)=p\Lambda^{\mathcal{M}}_{(1)}(t+\epsilon,t)+(1-p)\Lambda^{\mathcal{M}}_{(2)}(t+\epsilon,t),\\
=\mathbb{I}+\epsilon[p\mathcal{L}_t^{(1)}+(1-p)\mathcal{L}_t^{(2)}]=\mathbb{I}+\epsilon\mathcal{L}_t,
\end{array}
\]
with $\mathcal{L}_t=p\mathcal{L}_t^{(1)}+(1-p)\mathcal{L}_t^{(2)}$ and $0\leq p\leq 1$. Clearly, $\mathcal{L}_t$ is also a Lindblad type generator with positive coefficients, which proves $\Lambda^{\mathcal{M}}(t+\epsilon,t)$ also belongs to the set of divisible Markovian maps. Then, if the Choi states corresponding to the operations $\Lambda^{\mathcal{M}}_{(1,2)}$ are denoted by $\mathcal{C}^{\mathcal{M}}_{(1,2)}$, then their convex combination $\sigma=p\mathcal{C}^{\mathcal{M}}_{(1)}+(1-p)\mathcal{C}^{\mathcal{M}}_{(2)}$, can be represented as 
\[\sigma=\mathbb{I}\otimes\Lambda^{\mathcal{M}}(t+\epsilon,t)|\psi\ket\bra\psi|.\]
Since $\Lambda^{\mathcal{M}}(t+\epsilon,t)$ is a divisible map, we have $\sigma\in\mathbb{F}$.
\qed

Further, it can also be shown that in the limit of $\epsilon\rightarrow 0$, the set of all Choi states $\mathcal{A}$ is also a convex set. Using \textbf{Proposition 2}, we define the following measure of NM, taking the right derivative of $M(t+\epsilon,t)$ in the CP breaking region as 
\beq\label{n1}
\mathcal{D}_T(t)=\lim_{\epsilon\rightarrow 0^+}\frac{M_T(t+\epsilon,t)-0}{\epsilon},
\eeq
where $M_T(t+\epsilon,t)=\inf_{\mathcal{C}^{\mathcal{M}}(t+\epsilon,t)}||(\mathcal{C}^{\mathcal{N}}(t+\epsilon,t)-\mathcal{C}^{\mathcal{M}}(t+\epsilon,t)||_1$, with $||\cdot||_1$ is the trace norm. The minimum distance $M$ at the previous time $t$ is taken to be zero, since divisibility was not broken before $t$, the Choi state at that time belongs to the set of free states. For any quantum evolution, we always have $\mathcal{D}_T(t)\geq 0$. The equality holds for divisible Markovian evolutions. The optimization involved in evaluating $\mathcal{D}_T(t)$ is now easier because the set of free states now forms a convex set. Moreover, by virtue of \textbf{Proposition 1B} and \textbf{Proposition 2}, we know that the set of free states $\mathbb{F}$ in the limit $\epsilon\rightarrow 0$ is convex and compact. We can thus apply the Krein-Milman
theorem \citep{M1} to state that $\mathbb{F}$ is the convex hull of its extreme points. Hereafter, for brevity we use the short notation $\mathcal{C}^{\mathcal{N}(\mathcal{M})}(t+\epsilon,t)=\mathcal{C}^{\mathcal{N}(\mathcal{M})}(t)$. 

\noindent\textbf{Proposition 3:} \textit{$\mathcal{D}_T(t)$ is a bona fide measure of NM.}

\proof To prove $\mathcal{D}_T(t)$ is a bona fide measure, we have to prove it is faithful, convex and a monotone under the free operations. 

The measure is faithful iff $\mathcal{D}_T(t) \geq 0~~\mbox{and}~~\mathcal{D}_T(t)=0\Leftrightarrow \mathcal{C}^{\mathcal{N}}(t) \in \mathbb{F}.$  Clearly $\mathcal{D}_{T} (\mathcal{C}^{\mathcal{N}}(t)) \geq 0$  $\forall \mathcal{C}^{\mathcal{N}}(t)$ from the definition.
We now show that $\mathcal{D}_{T} (\mathcal{C}^{\mathcal{N}}(t)) = 0$ iff $\mathcal{C}^{\mathcal{N}}(t) \in \mathbb{F}$.
Here the if part is obvious from the definition. To prove the only if part,
let $\mathcal{D}_{T} (\mathcal{C}^{\mathcal{N}}(t)) = 0$ for some $\mathcal{C}^{\mathcal{N}}(t)$. Then
$\mathcal{D}_{T} (\mathcal{C}^{\mathcal{N}}(t)) = 0$
$\Rightarrow lim_{\epsilon \rightarrow 0} \dfrac{M(\mathcal{C}^{\mathcal{N}}(t))}{\epsilon}$ = 0
$\Rightarrow M_T(\mathcal{C}^{\mathcal{N}}(t)) = 0$, since $\epsilon$ is a finite positive number. Therefore
$ M_T(\mathcal{C}^{\mathcal{N}}(t))\Rightarrow inf_{\mathcal{C}^{\mathcal{M}}(t) \in \mathbb{F}} D(\mathcal{C}^{\mathcal{N}}(t) \vert \mathcal{C}^{\mathcal{M}}(t)) = 0$. This implies
$\mathcal{C}^{\mathcal{N}}(t) \in Cl(\mathbb{F}) = \mathbb{F}$ since $\mathbb{F}$ is a closed set. Therefore $\mathcal{D}_T(t)$ is faithful.

To prove the convexity of $\mathcal{D}_T(t)$, we consider $\mathcal{C}^{\mathcal{N}}_{1}(t)$ and $\mathcal{C}^{\mathcal{N}}_{2}(t)$ be two Choi states. By the convexity property of the set of all Choi states $\mathcal{A}$, we know $\mathcal{C}^{\mathcal{N}}(t)=p\mathcal{C}^{\mathcal{N}}_{1}(t)+(1-p)\mathcal{C}^{\mathcal{N}}_{2}(t)$ is also a Choi state. Therefore, by virtue of triangle inequality, we have $|| p\mathcal{C}^{\mathcal{N}}_1(t)+(1-p)\mathcal{C}^{\mathcal{N}}_{2}(t)-\mathcal{C}^{\mathcal{M}}(t)||_1 \leq p||\mathcal{C}^{\mathcal{N}}_1(t)-\mathcal{C}^{\mathcal{M}}(t)||_1+(1-p)||\mathcal{C}^{\mathcal{N}}_2(t)-\mathcal{C}^{\mathcal{M}}(t)||_1$, for all $\mathcal{C}^{\mathcal{M}}(t)$. Consequently, we get $\inf_{\mathcal{C}^{\mathcal{M}}(t)}|| p\mathcal{C}^{\mathcal{N}}_1(t)+(1-p)\mathcal{C}^{\mathcal{N}}_{2}(t)-\mathcal{C}^{\mathcal{M}}(t)||_1\leq p\inf_{\mathcal{C}^{\mathcal{M}}(t)}||\mathcal{C}^{\mathcal{N}}_1(t)-\mathcal{C}^{\mathcal{M}}(t)||_1+(1-p)\inf_{\mathcal{
 C}^{\mathcal{M}}(t)}||\mathcal{C}^{\mathcal{N}}_2(t)-\mathcal{C}^{\mathcal{M}}(t)||_1$. This in turn proves the convexity relation $\mathcal{D}_T(p\mathcal{C}^{\mathcal{M}}_1(t)+(1-p)\mathcal{C}^{\mathcal{M}}_2(t))\leq p\mathcal{D}_T(\mathcal{C}^{\mathcal{M}}_1(t))+(1-p)\mathcal{D}_T(\mathcal{C}^{\mathcal{M}}_2(t))$.

To prove the monotonicity of $\mathcal{D}_T(t)$, we consider a divisible free operation: $\rho(t_2)=\mathbb{I}\otimes\Lambda(t_2,t_1)(\rho(t_1))=Tr_E\left[V(t_2,t_1)\rho(t_1)\otimes\sigma_E V^{\dagger}(t_2,t_1)\right]$, where $V(t)$ is a global unitary acting on the composite system-environment Hilbert space and $\sigma_E$ is the initial state of the environment. Therefore, we have
$||\mathcal{C}^{\mathcal{N}}(t+\Delta))-\mathcal{C}^{\mathcal{M}}(t+\Delta)||_1=||Tr_E\left[V(t+\Delta,t)(\mathcal{C}^{\mathcal{N}}(t))-\mathcal{C}^{\mathcal{M}}(t))\otimes\sigma_E V^{\dagger}(t+\Delta,t)\right]||_1$. Using the trace norm inequality $||Tr_b[A_{ab}]||_1\leq ||[A_{ab}]||_1$, for any bounded operator $A_{ab}$, and preservation of trace norm under unitary rotation, we thus have 
$||(\mathcal{C}^{\mathcal{N}}(t))-
\mathcal{C}^{\mathcal{M}}(t))||_1=||{(\mathcal{C}^{\mathcal{N}}(t)-\mathcal{C}^{\mathcal{M}}(t))})\otimes\sigma_E||_1=||V(t+\Delta,t)(\mathcal{C}^{\mathcal{N}}(t))-\mathcal{C}^{\mathcal{M}}(t))\otimes\sigma_E V^{\dagger}(t+\Delta,t)||_1 \leq ||Tr_E[V(t+\Delta,t)(\mathcal{C}^{\mathcal{N}}(t))-\mathcal{C}^{\mathcal{M}}(t))\otimes\sigma_E V^{\dagger}(t+\Delta,t)]||_1=||(\mathcal{C}^{\mathcal{N}}(t+\Delta)-\mathcal{C}^{\mathcal{M}}(t+\Delta)))||_1$. 
Therefore we have $||(\mathcal{C}^{\mathcal{N}}(t+\Delta)-\mathcal{C}^{\mathcal{M}}(t+\Delta)))||_1 \leq ||(\mathcal{C}^{\mathcal{N}}(t)-\mathcal{C}^{\mathcal{M}}(t)))||_1$. 
 
 Now $M_T(t',t)=\inf_{\mathcal{C}^{\mathcal{M}}(t)}||\mathcal{C}^{\mathcal{N}}(t)-\mathcal{C}^{\mathcal{M}}(t)||_1=||\mathcal{C}^{\mathcal{N}}(t)-\mathcal{C}^{\mathcal{M*}}(t)||_1$, with $\mathcal{C}^{\mathcal{M*}}(t)$ being the free state from which the distance is minimum. Using the fact $\mathbb{I}\otimes\Lambda(t+\Delta,t)(\mathcal{C}^{\mathcal{M*}}(t)\in\mathbb{F}$, we have $\inf_{\mathbb{I}\otimes\Lambda(t+\Delta,t)(\mathcal{C}^{\mathcal{M*}}(t)}||\mathbb{I}\otimes\Lambda(t+\Delta,t)(\mathcal{C}^{\mathcal{N}}(t))-\mathbb{I}\otimes\Lambda(t+\Delta,t)(\mathcal{C}^{\mathcal{M*}}(t))||_1\leq ||\mathbb{I}\otimes\Lambda(t+\Delta,t)(\mathcal{C}^{\mathcal{N}}(t))-\mathbb{I}\otimes\Lambda(t+\Delta,t)(\mathcal{C}^{\mathcal{M*}}(t))||_1\leq||\mathcal{C}^{\mathcal{N}}(t)-\mathcal{C}^{\mathcal{M*}}(t)||_1$. It is evident from Eq. \eqref{n1} that $\mathcal{D}_T(t+\Delta)\leq \mathcal{D}_T(t)$, proving the monotonicity of $\
 mathcal{D}_T(t)$ under divisible operations. This completes the proof of the proposition. \qed

We further reduce the complexity of calculating the measure of NM, by constructing a lower bound of $\mathcal{D}_T(t)$ in the following theorem.

\noindent\textbf{Theorem 1:} 
\textit{Let $\Lambda^{\mathcal{N}}$ be map corresponding to some operation $\mathcal{N}$ and $g^{\mathcal{N}}(t)$ be the RHP measure, then $\mathcal{D}_T(t)$ is bounded below by $g^{\mathcal{N}}(t)$, i.e. $\mathcal{D}_T(t)\geq g^{\mathcal{N}}(t)$.}
\proof We have the expression of our NM measure
\[
\mathcal{D}_T(t)=\lim_{\epsilon\rightarrow 0^+}\frac{\inf_{\mathcal{C}^{\mathcal{M}}}\parallel\mathcal{C}^{\mathcal{N}}(t)-\mathcal{C}^{\mathcal{M}}(t)\parallel_1}{\epsilon}.
\]
Using the reverse triangle inequality: $||A-B||_1 \geq |\parallel A\parallel_1-\parallel B\parallel_1|$ with $\parallel\mathcal{C}^{\mathcal{N}}(t)\parallel_1\geq 1$ and $\parallel\mathcal{C}^{\mathcal{M}}(t)\parallel_1=1~~\forall \Lambda_{\mathcal{M}}$, we have
\[\mathcal{D}_T(t)\geq \lim_{\epsilon\rightarrow 0^+}\frac{\parallel\mathcal{C}^{\mathcal{N}}(t)\parallel_1-1}{\epsilon}=g^{\mathcal{N}}(t)\] \qed 

Interestingly, $\mathcal{D}_T(t)$ is lower bounded by  $g^{\mathcal{N}}(t)$, which is optimization free and easier to calculate. Note that $g^{\mathcal{N}}(t)$  is the time derivative of the trace norm of the Choi state \citep{rhp1}. When the divisibility breaks down, the norm of the corresponding Choi state is strictly greater that $1$. In those regions we have $g^{\mathcal{N}}(t)>1$, showing that the RHP measure is a witness of CP-indivisibility, whereas $\mathcal{D}_T(t)$ is the time derivative of the minimum distance between the Choi states corresponding to a specific evolution and all possible divisible maps. This, with the above mentioned lower bound proves that $\mathcal{D}_T(t)$ is a more general measure of NM. 

\noindent\textbf{Robustness of non-Markovianity:} Finally, we construct the concept of robustness of NM (RONM), in the similar footings of entanglement \citep{rob1,rob2,rob3}, coherence \citep{rob4} and asymmetry \citep{rob5}. In accordance with the definitions of robustness for other quantum resources, we define RONM as the minimum amount of Markovian noise needed to be added to a NM evolution to make the resulting evolution Markovian. Hence, the formal definition of RONM follows as

\beq\label{rob1}
\mathcal{R}_{\mathcal{N}}(\mathcal{C}^{\mathcal{N}}(t))=\inf_s\left\{s\geq 0: \frac{\mathcal{C}^{\mathcal{N}}(t)+s\tau^{\mathcal{N}}(t)}{1+s}=\delta^{\mathcal{M}}(t) \in \mathbb{F} \right\},
\eeq
where $\tau^{\mathcal{N}}(t)$ is an arbitrary element from the set of all Choi states $\mathcal{A}$.  
After achieving the minimization for Choi states $\tau^{\mathcal{N}*(t)}$ and $\delta^{\mathcal{M}*(t)}$, we write 
\beq\label{rob2}
\mathcal{C}^{\mathcal{N}}(t)=[1+\mathcal{R}_{\mathcal{N}}(\mathcal{C}^{\mathcal{N}}(t))]\delta^{\mathcal{M}*}(t)-\mathcal{R}_{\mathcal{N}}(\mathcal{C}^{\mathcal{N}}(t))\tau^{\mathcal{N}*}(t).
\eeq
In the following propositions, we establish that $\mathcal{R}_{\mathcal{N}}(\mathcal{C}^{\mathcal{N}}(t))$ is a bona fide measure of NM.

\noindent\textbf{Proposition 4:} \textit{$\mathcal{R}_{\mathcal{N}}(\mathcal{C}^{\mathcal{N}}(t))$ is a faithful, convex measure of NM, contractive under divisible operations.}

\proof The faithfulness of RONM follows from the definition as 
\[  
\mathcal{R}_{\mathcal{N}}(\mathcal{C}^{\mathcal{N}}(t)) \geq 0~~\mbox{and}~~\mathcal{R}_{\mathcal{N}}(\mathcal{C}^{\mathcal{N}}(t))=0\Leftrightarrow \mathcal{R}_{\mathcal{N}}(\mathcal{C}^{\mathcal{N}}(t)) \in \mathbb{F}.
\]
To prove the convexity of $\mathcal{R}_{\mathcal{N}}(\mathcal{C}^{\mathcal{N}}(t))$, 
let us consider two arbitrary Choi states $\mathcal{C}^{\mathcal{N}}_1(t)$ and $\mathcal{C}^{\mathcal{N}}_2(t)$, 
expressed as the pseudo-mixture $\mathcal{C}^{\mathcal{N}}_l(t)=[1+\mathcal{R}_{\mathcal{N}}(\mathcal{C}^{\mathcal{N}}_l(t))]\delta^{\mathcal{M}*}_l(t)-\mathcal{R}_{\mathcal{N}}(\mathcal{C}^{\mathcal{N}}_l(t))\tau^{\mathcal{N}*}_l(t)$ (for $l=1,2$). 
The convex structure of $\mathcal{A}$ ensures that the convex combination of these two Choi states will also be another Choi state. Now considering the convex decomposition $\mathcal{C}^{\mathcal{N}}(t)=p\mathcal{C}^{\mathcal{N}}_1(t)+(1-p)\mathcal{C}^{\mathcal{N}}_2(t)$ (with $0\leq p \leq 1$) 
and utilizing the pseudo-mixtures written above, the following pseudo-mixture $\mathcal{C}^{\mathcal{N}}(t)=[1+s]\delta^{\mathcal{M}}(t)-s(\mathcal{C}^{\mathcal{N}}(t))\tau^{\mathcal{N}}(t)$ 
can be written with $\delta^{\mathcal{M}}(t)=\{p[1+\mathcal{R}_{\mathcal{N}}(\mathcal{C}^{\mathcal{N}}_l(t))]\delta^{\mathcal{M}*}_l(t)+(1-p)[1+\mathcal{R}_{\mathcal{N}}(\mathcal{C}^{\mathcal{N}}_2(t))]\delta^{\mathcal{M}*}_2(t)\}/(1+s)$ and  
 $\tau^{\mathcal{N}}(t)=\{p\mathcal{R}_{\mathcal{N}}(\mathcal{C}^{\mathcal{N}}_l(t))\tau^{\mathcal{N}*}_l(t)+(1-p)\mathcal{R}_{\mathcal{N}}(\mathcal{C}^{\mathcal{N}}_2(t))\tau^{\mathcal{N}*}_2(t) \}/s$. The trace preservation property of any arbitrary quantum operation guarantees the hermiticity of Choi matrix with unit trace. Therefore, from the normalization condition, we get $s=p\mathcal{R}_{\mathcal{N}}(\mathcal{C}^{\mathcal{N}}_1(t))+(1-p)\mathcal{R}_{\mathcal{N}}(\mathcal{C}^{\mathcal{N}}_2(t))$. Now, from the definition of RONM, we have $\mathcal{R}_{\mathcal{N}}(\mathcal{C}^{\mathcal{N}}(t))\leq s$. Thus, the convexity of $\mathcal{R}_{\mathcal{N}}(\mathcal{C}^{\mathcal{N}}(t))$ is proved. 

To prove the monotonicity of $\mathcal{R}_{\mathcal{N}}(\mathcal{C}^{\mathcal{N}}(t))$, we use a result from entanglement theory \citep{rob6}. As we have seen from Eq. \eqref{rob2}, the Choi state $\mathcal{C}^{\mathcal{N}}(t)$ can be written in the pseudo mixture $\mathcal{C}^{\mathcal{N}}(t)=c_{+}\delta^{\mathcal{M}*}-c_{-}\tau^{\mathcal{M}*}$, with $c_+=1+\mathcal{R}_{\mathcal{N}}(\mathcal{C}^{\mathcal{N}}(t))$ and $c_-=\mathcal{R}_{\mathcal{N}}(\mathcal{C}^{\mathcal{N}}(t))$. If $\mathcal{P}_-$ be the projector onto the negative eigenvalue subspace of $\mathcal{C}^{\mathcal{N}}(t)$, then $c_-=-Tr[\mathcal{P}_-\mathcal{C}^{\mathcal{N}}(t)]$ is the sum of absolute values of negative eigenvalues. Let us consider the spectral decomposition $\mathcal{C}^{\mathcal{N}}(t)=\sum_{k}\lambda_k^+|\lambda_k^+\ket+\sum_{l}\lambda_l^-|\lambda_l^-\ket$ with $\lambda_k^+$s and $\lambda_l^-$s being the positive and negative eigenvalues respectively. 
 Therefore, the trace preservation condition yields $\sum_{k}\lambda_k^+ = 1+\sum_{l}|\lambda_l^-|$. Thus, we have $||\mathcal{C}^{\mathcal{N}}(t)||_1=\sum_k\lambda_k^+ + \sum_l|\lambda_l^-|=1+2\sum_l|\lambda_l^-|$. Since $\mathcal{R}_{\mathcal{N}}(\mathcal{C}^{\mathcal{N}}(t))=c_-=\sum_l|\lambda_l^-|$, we get the following relation
\beq\label{robN}
\mathcal{R}_{\mathcal{N}}(\mathcal{C}^{\mathcal{N}}(t))=\frac{||\mathcal{C}^{\mathcal{N}}(t)||_1-1}{2}.
\eeq
From the proof of \textbf{Proposition 3}, we know that trace norm is contractive under divisible operations. Hence, from Eq. \eqref{robN}, it is evident that $\mathcal{R}_{\mathcal{N}}(\mathcal{C}^{\mathcal{N}}(t))$ is contractive under divisible operations.
\qed

By virtue of \textbf{Proposition 4}, we surmise that $\mathcal{R}_{\mathcal{N}}(\mathcal{C}^{\mathcal{N}}(t))$ is also a bona fide measure of NM. If $\mathcal{C}^{\mathcal{N}}_1(t)$ and $\mathcal{C}^{\mathcal{N}}_2(t)$ be two arbitrary Choi states, then their convex mixture can be written as $(\mathcal{C}^{\mathcal{N}}_1(t)+s\mathcal{C}^{\mathcal{N}}_1(t))/(1+s)=[\mathbb{I}\otimes\mathbb{I}+\mathbb{I}\otimes(1/(1+s)\mathcal{L}_t^{(1)}+s/(1+s)\mathcal{L}_t^{(2)})]|\psi\ket\bra\psi|$.  RONM physically means the endurance of a NM operation under mixing with Markovian noise. From Eq. \eqref{robN}, we see that there is an optimization free way to compute $\mathcal{R}_{\mathcal{N}}(\mathcal{C}^{\mathcal{N}}(t))$, and hence,  we get a bona fide measure of NM which is easy to compute. There is another important conclusion that can be drawn from the proof of \textbf{Proposition 4}. From Eq. \eqref{robN} we get that the RONM can be expressed in terms of the RHP measure of NM  as
\beq\label{ronN1}
\mathcal{R}_{\mathcal{N}}(\mathcal{C}^{\mathcal{N}}(t))=\frac{1}{2}\mathcal{N}_T(t)=\frac{1}{2}\int_0^t g^{\mathcal{N}}(t')dt'.
\eeq 
So the RONM can be directly expressed as a function of the RHP measure $g^{\mathcal{N}}(t)$. This provides a physical interpretation of the RHP measure. The  normalized measure of NM \citep{rhp1} can therefore be expressed as
\[\mathcal{T}(t)=\frac{\mathcal{N}_T(t)}{1+\mathcal{N}_T(t)}=\frac{2\mathcal{R}_{\mathcal{N}}(\mathcal{C}^{\mathcal{N}}(t))}{1+2\mathcal{R}_{\mathcal{N}}(\mathcal{C}^{\mathcal{N}}(t))}.\]

\textit{Conclusion:} In this work, we have constructed a convex RTNM under the constraint of small time interval, which satisfies all the properties of RT. We have defined the divisible operations as the free operations, Choi states corresponding to the divisible operations as the free states and constructed a bona fide measure of NM. In a recent work \citep{modi}, a  measure of  NM in terms of minimum quasi-distances has been proposed, where the minimization is done over all Markov processes. Due to the non-convexity of Markov processes, this optimization is extremely difficult to compute in practice. On the other hand, calculating our proposed measure $\mathcal{D}_T(t)$ is much easier, since the free Choi states ($\mathcal{C}^{\mathcal{M}}(t)$) form a convex set in a sufficiently small time interval. Moreover, $\mathcal{D}_T(t)$ is lower bounded by the  optimization free RHP measure of NM. We have also constructed robustness of NM, which is the 
 degree of endurance of any NM operation under the mixing with Markovian noise. We have shown that it satisfies all the properties of a bona fide measure, and is also easy to compute from the optimization free relation \eqref{robN}. Moreover, we have directly connected RONM with RHP measure , which allows us to present an operational interpretation of RHP measure of NM.  
Applications of the present results in future work may ascertain the practical importance of our RTNM in the study of NM in quantum information and thermodynamics. \\

SB thanks Sibasish Ghosh of Institute of Mathematical Sciences and Manik Banik of S.N. Bose National Centre for Basic Sciences for fruitful discussion. SB acknowledges SERB, DST, Government of India for financial support. BB acknowledges DST INSPIRE programme for financial support.

\bibliographystyle{apsrev4-1}
\bibliography{Athermality1}

\begin{thebibliography}{54}%
\makeatletter
\providecommand \@ifxundefined [1]{%
 \@ifx{#1\undefined}
}%
\providecommand \@ifnum [1]{%
 \ifnum #1\expandafter \@firstoftwo
 \else \expandafter \@secondoftwo
 \fi
}%
\providecommand \@ifx [1]{%
 \ifx #1\expandafter \@firstoftwo
 \else \expandafter \@secondoftwo
 \fi
}%
\providecommand \natexlab [1]{#1}%
\providecommand \enquote  [1]{``#1''}%
\providecommand \bibnamefont  [1]{#1}%
\providecommand \bibfnamefont [1]{#1}%
\providecommand \citenamefont [1]{#1}%
\providecommand \href@noop [0]{\@secondoftwo}%
\providecommand \href [0]{\begingroup \@sanitize@url \@href}%
\providecommand \@href[1]{\@@startlink{#1}\@@href}%
\providecommand \@@href[1]{\endgroup#1\@@endlink}%
\providecommand \@sanitize@url [0]{\catcode `\\12\catcode `\$12\catcode
  `\&12\catcode `\#12\catcode `\^12\catcode `\_12\catcode `\%12\relax}%
\providecommand \@@startlink[1]{}%
\providecommand \@@endlink[0]{}%
\providecommand \url  [0]{\begingroup\@sanitize@url \@url }%
\providecommand \@url [1]{\endgroup\@href {#1}{\urlprefix }}%
\providecommand \urlprefix  [0]{URL }%
\providecommand \Eprint [0]{\href }%
\providecommand \doibase [0]{http://dx.doi.org/}%
\providecommand \selectlanguage [0]{\@gobble}%
\providecommand \bibinfo  [0]{\@secondoftwo}%
\providecommand \bibfield  [0]{\@secondoftwo}%
\providecommand \translation [1]{[#1]}%
\providecommand \BibitemOpen [0]{}%
\providecommand \bibitemStop [0]{}%
\providecommand \bibitemNoStop [0]{.\EOS\space}%
\providecommand \EOS [0]{\spacefactor3000\relax}%
\providecommand \BibitemShut  [1]{\csname bibitem#1\endcsname}%
\let\auto@bib@innerbib\@empty
\bibitem [{\citenamefont {Alicki}\ and\ \citenamefont {Lendi}(2007)}]{alicki}%
  \BibitemOpen
  \bibfield  {author} {\bibinfo {author} {\bibfnamefont {R.}~\bibnamefont
  {Alicki}}\ and\ \bibinfo {author} {\bibfnamefont {K.}~\bibnamefont {Lendi}},\
  }\href@noop {} {\emph {\bibinfo {title} {Quantum Dynamical Semigroups and
  Applications}}},\ Lecture notes in Physics\ (\bibinfo  {publisher}
  {Springer-Verlag Berlin Heidelberg},\ \bibinfo {year} {2007})\BibitemShut
  {NoStop}%
\bibitem [{\citenamefont {{Lindblad}}(1976)}]{lindblad}%
  \BibitemOpen
  \bibfield  {author} {\bibinfo {author} {\bibfnamefont {G.}~\bibnamefont
  {{Lindblad}}},\ }\href {\doibase 10.1007/BF01608499} {\bibfield  {journal}
  {\bibinfo  {journal} {Communications in Mathematical Physics}\ }\textbf
  {\bibinfo {volume} {48}},\ \bibinfo {pages} {119} (\bibinfo {year}
  {1976})}\BibitemShut {NoStop}%
\bibitem [{\citenamefont {{Gorini}}\ \emph {et~al.}(1976)\citenamefont
  {{Gorini}}, \citenamefont {{Kossakowski}},\ and\ \citenamefont
  {{Sudarshan}}}]{gorini}%
  \BibitemOpen
  \bibfield  {author} {\bibinfo {author} {\bibfnamefont {V.}~\bibnamefont
  {{Gorini}}}, \bibinfo {author} {\bibfnamefont {A.}~\bibnamefont
  {{Kossakowski}}}, \ and\ \bibinfo {author} {\bibfnamefont {E.~C.~G.}\
  \bibnamefont {{Sudarshan}}},\ }\href {\doibase 10.1063/1.522979} {\bibfield
  {journal} {\bibinfo  {journal} {Journal of Mathematical Physics}\ }\textbf
  {\bibinfo {volume} {17}},\ \bibinfo {pages} {821} (\bibinfo {year}
  {1976})}\BibitemShut {NoStop}%
\bibitem [{\citenamefont {Breuer}\ and\ \citenamefont
  {Petruccione}(2002)}]{breuer}%
  \BibitemOpen
  \bibfield  {author} {\bibinfo {author} {\bibfnamefont {H.~P.}\ \bibnamefont
  {Breuer}}\ and\ \bibinfo {author} {\bibfnamefont {F.}~\bibnamefont
  {Petruccione}},\ }\href@noop {} {\emph {\bibinfo {title} {The theory of open
  quantum systems}}}\ (\bibinfo  {publisher} {Oxford University Press},\
  \bibinfo {address} {Great Clarendon Street},\ \bibinfo {year}
  {2002})\BibitemShut {NoStop}%
\bibitem [{\citenamefont {Rivas}\ \emph {et~al.}(2014)\citenamefont {Rivas},
  \citenamefont {Huelga},\ and\ \citenamefont {Plenio}}]{rivas1}%
  \BibitemOpen
  \bibfield  {author} {\bibinfo {author} {\bibfnamefont {A.}~\bibnamefont
  {Rivas}}, \bibinfo {author} {\bibfnamefont {S.~F.}\ \bibnamefont {Huelga}}, \
  and\ \bibinfo {author} {\bibfnamefont {M.~B.}\ \bibnamefont {Plenio}},\
  }\href {http://stacks.iop.org/0034-4885/77/i=9/a=094001} {\bibfield
  {journal} {\bibinfo  {journal} {Reports on Progress in Physics}\ }\textbf
  {\bibinfo {volume} {77}},\ \bibinfo {pages} {094001} (\bibinfo {year}
  {2014})}\BibitemShut {NoStop}%
\bibitem [{\citenamefont {Breuer}\ \emph {et~al.}(2016)\citenamefont {Breuer},
  \citenamefont {Laine}, \citenamefont {Piilo},\ and\ \citenamefont
  {Vacchini}}]{breuerN}%
  \BibitemOpen
  \bibfield  {author} {\bibinfo {author} {\bibfnamefont {H.-P.}\ \bibnamefont
  {Breuer}}, \bibinfo {author} {\bibfnamefont {E.-M.}\ \bibnamefont {Laine}},
  \bibinfo {author} {\bibfnamefont {J.}~\bibnamefont {Piilo}}, \ and\ \bibinfo
  {author} {\bibfnamefont {B.}~\bibnamefont {Vacchini}},\ }\href {\doibase
  10.1103/RevModPhys.88.021002} {\bibfield  {journal} {\bibinfo  {journal}
  {Rev. Mod. Phys.}\ }\textbf {\bibinfo {volume} {88}},\ \bibinfo {pages}
  {021002} (\bibinfo {year} {2016})}\BibitemShut {NoStop}%
\bibitem [{\citenamefont {de~Vega}\ and\ \citenamefont
  {Alonso}(2017)}]{alonso}%
  \BibitemOpen
  \bibfield  {author} {\bibinfo {author} {\bibfnamefont {I.}~\bibnamefont
  {de~Vega}}\ and\ \bibinfo {author} {\bibfnamefont {D.}~\bibnamefont
  {Alonso}},\ }\href {\doibase 10.1103/RevModPhys.89.015001} {\bibfield
  {journal} {\bibinfo  {journal} {Rev. Mod. Phys.}\ }\textbf {\bibinfo {volume}
  {89}},\ \bibinfo {pages} {015001} (\bibinfo {year} {2017})}\BibitemShut
  {NoStop}%
\bibitem [{\citenamefont {Rivas}\ \emph {et~al.}(2010)\citenamefont {Rivas},
  \citenamefont {Huelga},\ and\ \citenamefont {Plenio}}]{rhp1}%
  \BibitemOpen
  \bibfield  {author} {\bibinfo {author} {\bibfnamefont {A.}~\bibnamefont
  {Rivas}}, \bibinfo {author} {\bibfnamefont {S.~F.}\ \bibnamefont {Huelga}}, \
  and\ \bibinfo {author} {\bibfnamefont {M.~B.}\ \bibnamefont {Plenio}},\
  }\href {\doibase 10.1103/PhysRevLett.105.050403} {\bibfield  {journal}
  {\bibinfo  {journal} {Phys. Rev. Lett.}\ }\textbf {\bibinfo {volume} {105}},\
  \bibinfo {pages} {050403} (\bibinfo {year} {2010})}\BibitemShut {NoStop}%
\bibitem [{\citenamefont {Laine}\ \emph {et~al.}(2010)\citenamefont {Laine},
  \citenamefont {Piilo},\ and\ \citenamefont {Breuer}}]{blp1}%
  \BibitemOpen
  \bibfield  {author} {\bibinfo {author} {\bibfnamefont {E.-M.}\ \bibnamefont
  {Laine}}, \bibinfo {author} {\bibfnamefont {J.}~\bibnamefont {Piilo}}, \ and\
  \bibinfo {author} {\bibfnamefont {H.-P.}\ \bibnamefont {Breuer}},\ }\href
  {\doibase 10.1103/PhysRevA.81.062115} {\bibfield  {journal} {\bibinfo
  {journal} {Phys. Rev. A}\ }\textbf {\bibinfo {volume} {81}},\ \bibinfo
  {pages} {062115} (\bibinfo {year} {2010})}\BibitemShut {NoStop}%
\bibitem [{\citenamefont {Chen}\ \emph {et~al.}(2018)\citenamefont {Chen},
  \citenamefont {Gneiting}, \citenamefont {Lo}, \citenamefont {Chen},\ and\
  \citenamefont {Nori}}]{proman1}%
  \BibitemOpen
  \bibfield  {author} {\bibinfo {author} {\bibfnamefont {H.-B.}\ \bibnamefont
  {Chen}}, \bibinfo {author} {\bibfnamefont {C.}~\bibnamefont {Gneiting}},
  \bibinfo {author} {\bibfnamefont {P.-Y.}\ \bibnamefont {Lo}}, \bibinfo
  {author} {\bibfnamefont {Y.-N.}\ \bibnamefont {Chen}}, \ and\ \bibinfo
  {author} {\bibfnamefont {F.}~\bibnamefont {Nori}},\ }\href {\doibase
  10.1103/PhysRevLett.120.030403} {\bibfield  {journal} {\bibinfo  {journal}
  {Phys. Rev. Lett.}\ }\textbf {\bibinfo {volume} {120}},\ \bibinfo {pages}
  {030403} (\bibinfo {year} {2018})}\BibitemShut {NoStop}%
\bibitem [{\citenamefont {Chanda}\ and\ \citenamefont
  {Bhattacharya}(2016)}]{chanda}%
  \BibitemOpen
  \bibfield  {author} {\bibinfo {author} {\bibfnamefont {T.}~\bibnamefont
  {Chanda}}\ and\ \bibinfo {author} {\bibfnamefont {S.}~\bibnamefont
  {Bhattacharya}},\ }\href {\doibase https://doi.org/10.1016/j.aop.2016.01.004}
  {\bibfield  {journal} {\bibinfo  {journal} {Annals of Physics}\ }\textbf
  {\bibinfo {volume} {366}},\ \bibinfo {pages} {1 } (\bibinfo {year}
  {2016})}\BibitemShut {NoStop}%
\bibitem [{\citenamefont {Yu}\ \emph {et~al.}(2018)\citenamefont {Yu},
  \citenamefont {Wang}, \citenamefont {Ke}, \citenamefont {Liu}, \citenamefont
  {Meng}, \citenamefont {Li}, \citenamefont {Zhang}, \citenamefont {Chen},
  \citenamefont {Tang}, \citenamefont {Li},\ and\ \citenamefont
  {Guo}}]{proman2}%
  \BibitemOpen
  \bibfield  {author} {\bibinfo {author} {\bibfnamefont {S.}~\bibnamefont
  {Yu}}, \bibinfo {author} {\bibfnamefont {Y.-T.}\ \bibnamefont {Wang}},
  \bibinfo {author} {\bibfnamefont {Z.-J.}\ \bibnamefont {Ke}}, \bibinfo
  {author} {\bibfnamefont {W.}~\bibnamefont {Liu}}, \bibinfo {author}
  {\bibfnamefont {Y.}~\bibnamefont {Meng}}, \bibinfo {author} {\bibfnamefont
  {Z.-P.}\ \bibnamefont {Li}}, \bibinfo {author} {\bibfnamefont {W.-H.}\
  \bibnamefont {Zhang}}, \bibinfo {author} {\bibfnamefont {G.}~\bibnamefont
  {Chen}}, \bibinfo {author} {\bibfnamefont {J.-S.}\ \bibnamefont {Tang}},
  \bibinfo {author} {\bibfnamefont {C.-F.}\ \bibnamefont {Li}}, \ and\ \bibinfo
  {author} {\bibfnamefont {G.-C.}\ \bibnamefont {Guo}},\ }\href {\doibase
  10.1103/PhysRevLett.120.060406} {\bibfield  {journal} {\bibinfo  {journal}
  {Phys. Rev. Lett.}\ }\textbf {\bibinfo {volume} {120}},\ \bibinfo {pages}
  {060406} (\bibinfo {year} {2018})}\BibitemShut {NoStop}%
\bibitem [{\citenamefont {Chru\ifmmode \acute{s}\else
  \'{s}\fi{}ci\ifmmode~\acute{n}\else \'{n}\fi{}ski}\ \emph
  {et~al.}(2017)\citenamefont {Chru\ifmmode \acute{s}\else
  \'{s}\fi{}ci\ifmmode~\acute{n}\else \'{n}\fi{}ski}, \citenamefont
  {Macchiavello},\ and\ \citenamefont {Maniscalco}}]{proman3}%
  \BibitemOpen
  \bibfield  {author} {\bibinfo {author} {\bibfnamefont {D.}~\bibnamefont
  {Chru\ifmmode \acute{s}\else \'{s}\fi{}ci\ifmmode~\acute{n}\else
  \'{n}\fi{}ski}}, \bibinfo {author} {\bibfnamefont {C.}~\bibnamefont
  {Macchiavello}}, \ and\ \bibinfo {author} {\bibfnamefont {S.}~\bibnamefont
  {Maniscalco}},\ }\href {\doibase 10.1103/PhysRevLett.118.080404} {\bibfield
  {journal} {\bibinfo  {journal} {Phys. Rev. Lett.}\ }\textbf {\bibinfo
  {volume} {118}},\ \bibinfo {pages} {080404} (\bibinfo {year}
  {2017})}\BibitemShut {NoStop}%
\bibitem [{\citenamefont {Bylicka}\ \emph {et~al.}(2017)\citenamefont
  {Bylicka}, \citenamefont {Johansson},\ and\ \citenamefont
  {Ac\'{\i}n}}]{proman4}%
  \BibitemOpen
  \bibfield  {author} {\bibinfo {author} {\bibfnamefont {B.}~\bibnamefont
  {Bylicka}}, \bibinfo {author} {\bibfnamefont {M.}~\bibnamefont {Johansson}},
  \ and\ \bibinfo {author} {\bibfnamefont {A.}~\bibnamefont {Ac\'{\i}n}},\
  }\href {\doibase 10.1103/PhysRevLett.118.120501} {\bibfield  {journal}
  {\bibinfo  {journal} {Phys. Rev. Lett.}\ }\textbf {\bibinfo {volume} {118}},\
  \bibinfo {pages} {120501} (\bibinfo {year} {2017})}\BibitemShut {NoStop}%
\bibitem [{\citenamefont {Liuzzo-Scorpo}\ \emph {et~al.}(2017)\citenamefont
  {Liuzzo-Scorpo}, \citenamefont {Roga}, \citenamefont {Souza}, \citenamefont
  {Bernardes},\ and\ \citenamefont {Adesso}}]{proman5}%
  \BibitemOpen
  \bibfield  {author} {\bibinfo {author} {\bibfnamefont {P.}~\bibnamefont
  {Liuzzo-Scorpo}}, \bibinfo {author} {\bibfnamefont {W.}~\bibnamefont {Roga}},
  \bibinfo {author} {\bibfnamefont {L.~A.~M.}\ \bibnamefont {Souza}}, \bibinfo
  {author} {\bibfnamefont {N.~K.}\ \bibnamefont {Bernardes}}, \ and\ \bibinfo
  {author} {\bibfnamefont {G.}~\bibnamefont {Adesso}},\ }\href {\doibase
  10.1103/PhysRevLett.118.050401} {\bibfield  {journal} {\bibinfo  {journal}
  {Phys. Rev. Lett.}\ }\textbf {\bibinfo {volume} {118}},\ \bibinfo {pages}
  {050401} (\bibinfo {year} {2017})}\BibitemShut {NoStop}%
\bibitem [{\citenamefont {Kawabata}\ \emph {et~al.}(2017)\citenamefont
  {Kawabata}, \citenamefont {Ashida},\ and\ \citenamefont {Ueda}}]{proman8}%
  \BibitemOpen
  \bibfield  {author} {\bibinfo {author} {\bibfnamefont {K.}~\bibnamefont
  {Kawabata}}, \bibinfo {author} {\bibfnamefont {Y.}~\bibnamefont {Ashida}}, \
  and\ \bibinfo {author} {\bibfnamefont {M.}~\bibnamefont {Ueda}},\ }\href
  {\doibase 10.1103/PhysRevLett.119.190401} {\bibfield  {journal} {\bibinfo
  {journal} {Phys. Rev. Lett.}\ }\textbf {\bibinfo {volume} {119}},\ \bibinfo
  {pages} {190401} (\bibinfo {year} {2017})}\BibitemShut {NoStop}%
\bibitem [{\citenamefont {Bae}\ and\ \citenamefont {Chru\ifmmode \acute{s}\else
  \'{s}\fi{}ci\ifmmode~\acute{n}\else \'{n}\fi{}ski}(2016)}]{proman9}%
  \BibitemOpen
  \bibfield  {author} {\bibinfo {author} {\bibfnamefont {J.}~\bibnamefont
  {Bae}}\ and\ \bibinfo {author} {\bibfnamefont {D.}~\bibnamefont {Chru\ifmmode
  \acute{s}\else \'{s}\fi{}ci\ifmmode~\acute{n}\else \'{n}\fi{}ski}},\ }\href
  {\doibase 10.1103/PhysRevLett.117.050403} {\bibfield  {journal} {\bibinfo
  {journal} {Phys. Rev. Lett.}\ }\textbf {\bibinfo {volume} {117}},\ \bibinfo
  {pages} {050403} (\bibinfo {year} {2016})}\BibitemShut {NoStop}%
\bibitem [{\citenamefont {Bellomo}\ \emph {et~al.}(2007)\citenamefont
  {Bellomo}, \citenamefont {Lo~Franco},\ and\ \citenamefont
  {Compagno}}]{bellomo}%
  \BibitemOpen
  \bibfield  {author} {\bibinfo {author} {\bibfnamefont {B.}~\bibnamefont
  {Bellomo}}, \bibinfo {author} {\bibfnamefont {R.}~\bibnamefont {Lo~Franco}},
  \ and\ \bibinfo {author} {\bibfnamefont {G.}~\bibnamefont {Compagno}},\
  }\href {\doibase 10.1103/PhysRevLett.99.160502} {\bibfield  {journal}
  {\bibinfo  {journal} {Phys. Rev. Lett.}\ }\textbf {\bibinfo {volume} {99}},\
  \bibinfo {pages} {160502} (\bibinfo {year} {2007})}\BibitemShut {NoStop}%
\bibitem [{\citenamefont {Dijkstra}\ and\ \citenamefont
  {Tanimura}(2010)}]{arend}%
  \BibitemOpen
  \bibfield  {author} {\bibinfo {author} {\bibfnamefont {A.~G.}\ \bibnamefont
  {Dijkstra}}\ and\ \bibinfo {author} {\bibfnamefont {Y.}~\bibnamefont
  {Tanimura}},\ }\href {\doibase 10.1103/PhysRevLett.104.250401} {\bibfield
  {journal} {\bibinfo  {journal} {Phys. Rev. Lett.}\ }\textbf {\bibinfo
  {volume} {104}},\ \bibinfo {pages} {250401} (\bibinfo {year}
  {2010})}\BibitemShut {NoStop}%
\bibitem [{\citenamefont {{Kumar}}\ \emph {et~al.}(2017)\citenamefont
  {{Kumar}}, \citenamefont {{Banerjee}}, \citenamefont {{Srikanth}},
  \citenamefont {{Jagadish}},\ and\ \citenamefont {{Petruccione}}}]{ban1}%
  \BibitemOpen
  \bibfield  {author} {\bibinfo {author} {\bibfnamefont {P.}~\bibnamefont
  {{Kumar}}}, \bibinfo {author} {\bibfnamefont {S.}~\bibnamefont {{Banerjee}}},
  \bibinfo {author} {\bibfnamefont {R.}~\bibnamefont {{Srikanth}}}, \bibinfo
  {author} {\bibfnamefont {V.}~\bibnamefont {{Jagadish}}}, \ and\ \bibinfo
  {author} {\bibfnamefont {F.}~\bibnamefont {{Petruccione}}},\ }\href@noop {}
  {\bibfield  {journal} {\bibinfo  {journal} {ArXiv e-prints}\ } (\bibinfo
  {year} {2017})},\ \Eprint {http://arxiv.org/abs/1711.03267} {arXiv:1711.03267
  [quant-ph]} \BibitemShut {NoStop}%
\bibitem [{\citenamefont {{Pradeep Kumar}}\ \emph {et~al.}(2018)\citenamefont
  {{Pradeep Kumar}}, \citenamefont {{Banerjee}},\ and\ \citenamefont
  {{Chandrashekar}}}]{ban2}%
  \BibitemOpen
  \bibfield  {author} {\bibinfo {author} {\bibfnamefont {N.}~\bibnamefont
  {{Pradeep Kumar}}}, \bibinfo {author} {\bibfnamefont {S.}~\bibnamefont
  {{Banerjee}}}, \ and\ \bibinfo {author} {\bibfnamefont {C.~M.}\ \bibnamefont
  {{Chandrashekar}}},\ }\href@noop {} {\bibfield  {journal} {\bibinfo
  {journal} {ArXiv e-prints}\ } (\bibinfo {year} {2018})},\ \Eprint
  {http://arxiv.org/abs/1802.05478} {arXiv:1802.05478 [quant-ph]} \BibitemShut
  {NoStop}%
\bibitem [{\citenamefont {Bhattacharya}\ \emph {et~al.}(2017)\citenamefont
  {Bhattacharya}, \citenamefont {Misra}, \citenamefont {Mukhopadhyay},\ and\
  \citenamefont {Pati}}]{samya1}%
  \BibitemOpen
  \bibfield  {author} {\bibinfo {author} {\bibfnamefont {S.}~\bibnamefont
  {Bhattacharya}}, \bibinfo {author} {\bibfnamefont {A.}~\bibnamefont {Misra}},
  \bibinfo {author} {\bibfnamefont {C.}~\bibnamefont {Mukhopadhyay}}, \ and\
  \bibinfo {author} {\bibfnamefont {A.~K.}\ \bibnamefont {Pati}},\ }\href
  {\doibase 10.1103/PhysRevA.95.012122} {\bibfield  {journal} {\bibinfo
  {journal} {Phys. Rev. A}\ }\textbf {\bibinfo {volume} {95}},\ \bibinfo
  {pages} {012122} (\bibinfo {year} {2017})}\BibitemShut {NoStop}%
\bibitem [{\citenamefont {Mukhopadhyay}\ \emph {et~al.}(2017)\citenamefont
  {Mukhopadhyay}, \citenamefont {Bhattacharya}, \citenamefont {Misra},\ and\
  \citenamefont {Pati}}]{samya2}%
  \BibitemOpen
  \bibfield  {author} {\bibinfo {author} {\bibfnamefont {C.}~\bibnamefont
  {Mukhopadhyay}}, \bibinfo {author} {\bibfnamefont {S.}~\bibnamefont
  {Bhattacharya}}, \bibinfo {author} {\bibfnamefont {A.}~\bibnamefont {Misra}},
  \ and\ \bibinfo {author} {\bibfnamefont {A.~K.}\ \bibnamefont {Pati}},\
  }\href {\doibase 10.1103/PhysRevA.96.052125} {\bibfield  {journal} {\bibinfo
  {journal} {Phys. Rev. A}\ }\textbf {\bibinfo {volume} {96}},\ \bibinfo
  {pages} {052125} (\bibinfo {year} {2017})}\BibitemShut {NoStop}%
\bibitem [{\citenamefont {Chen}\ \emph {et~al.}(2016)\citenamefont {Chen},
  \citenamefont {Lambert}, \citenamefont {Li}, \citenamefont {Miranowicz},
  \citenamefont {Chen},\ and\ \citenamefont {Nori}}]{proman10}%
  \BibitemOpen
  \bibfield  {author} {\bibinfo {author} {\bibfnamefont {S.-L.}\ \bibnamefont
  {Chen}}, \bibinfo {author} {\bibfnamefont {N.}~\bibnamefont {Lambert}},
  \bibinfo {author} {\bibfnamefont {C.-M.}\ \bibnamefont {Li}}, \bibinfo
  {author} {\bibfnamefont {A.}~\bibnamefont {Miranowicz}}, \bibinfo {author}
  {\bibfnamefont {Y.-N.}\ \bibnamefont {Chen}}, \ and\ \bibinfo {author}
  {\bibfnamefont {F.}~\bibnamefont {Nori}},\ }\href {\doibase
  10.1103/PhysRevLett.116.020503} {\bibfield  {journal} {\bibinfo  {journal}
  {Phys. Rev. Lett.}\ }\textbf {\bibinfo {volume} {116}},\ \bibinfo {pages}
  {020503} (\bibinfo {year} {2016})}\BibitemShut {NoStop}%
\bibitem [{\citenamefont {Laine}\ \emph {et~al.}(2014)\citenamefont {Laine},
  \citenamefont {Breuer},\ and\ \citenamefont {Piilo}}]{task1}%
  \BibitemOpen
  \bibfield  {author} {\bibinfo {author} {\bibfnamefont {E.-M.}\ \bibnamefont
  {Laine}}, \bibinfo {author} {\bibfnamefont {H.-P.}\ \bibnamefont {Breuer}}, \
  and\ \bibinfo {author} {\bibfnamefont {J.}~\bibnamefont {Piilo}},\ }\href
  {https://www.nature.com/articles/srep04620} {\bibfield  {journal} {\bibinfo
  {journal} {Scientific Reports}\ }\textbf {\bibinfo {volume} {4}},\ \bibinfo
  {pages} {4620} (\bibinfo {year} {2014})}\BibitemShut {NoStop}%
\bibitem [{\citenamefont {Bylicka}\ \emph {et~al.}(2014)\citenamefont
  {Bylicka}, \citenamefont {Chruściński},\ and\ \citenamefont
  {Maniscalco}}]{task2}%
  \BibitemOpen
  \bibfield  {author} {\bibinfo {author} {\bibfnamefont {B.}~\bibnamefont
  {Bylicka}}, \bibinfo {author} {\bibfnamefont {D.}~\bibnamefont
  {Chruściński}}, \ and\ \bibinfo {author} {\bibfnamefont {S.}~\bibnamefont
  {Maniscalco}},\ }\href {https://www.nature.com/articles/srep05720} {\bibfield
   {journal} {\bibinfo  {journal} {Scientific Reports}\ }\textbf {\bibinfo
  {volume} {4}},\ \bibinfo {pages} {5720} (\bibinfo {year} {2014})}\BibitemShut
  {NoStop}%
\bibitem [{\citenamefont {Xiang}\ \emph {et~al.}(2014)\citenamefont {Xiang},
  \citenamefont {Hou}, \citenamefont {Li}, \citenamefont {Guo}, \citenamefont
  {Breuer}, \citenamefont {Laine},\ and\ \citenamefont {Piilo}}]{task3}%
  \BibitemOpen
  \bibfield  {author} {\bibinfo {author} {\bibfnamefont {G.-Y.}\ \bibnamefont
  {Xiang}}, \bibinfo {author} {\bibfnamefont {Z.-B.}\ \bibnamefont {Hou}},
  \bibinfo {author} {\bibfnamefont {C.-F.}\ \bibnamefont {Li}}, \bibinfo
  {author} {\bibfnamefont {G.-C.}\ \bibnamefont {Guo}}, \bibinfo {author}
  {\bibfnamefont {H.-P.}\ \bibnamefont {Breuer}}, \bibinfo {author}
  {\bibfnamefont {E.-M.}\ \bibnamefont {Laine}}, \ and\ \bibinfo {author}
  {\bibfnamefont {J.}~\bibnamefont {Piilo}},\ }\href
  {http://stacks.iop.org/0295-5075/107/i=5/a=54006} {\bibfield  {journal}
  {\bibinfo  {journal} {EPL (Europhysics Letters)}\ }\textbf {\bibinfo {volume}
  {107}},\ \bibinfo {pages} {54006} (\bibinfo {year} {2014})}\BibitemShut
  {NoStop}%
\bibitem [{\citenamefont {Thomas}\ \emph {et~al.}(2018)\citenamefont {Thomas},
  \citenamefont {Siddharth}, \citenamefont {Banerjee},\ and\ \citenamefont
  {Ghosh}}]{task4}%
  \BibitemOpen
  \bibfield  {author} {\bibinfo {author} {\bibfnamefont {G.}~\bibnamefont
  {Thomas}}, \bibinfo {author} {\bibfnamefont {N.}~\bibnamefont {Siddharth}},
  \bibinfo {author} {\bibfnamefont {S.}~\bibnamefont {Banerjee}}, \ and\
  \bibinfo {author} {\bibfnamefont {S.}~\bibnamefont {Ghosh}},\ }\href
  {\doibase 10.1103/PhysRevE.97.062108} {\bibfield  {journal} {\bibinfo
  {journal} {Phys. Rev. E}\ }\textbf {\bibinfo {volume} {97}},\ \bibinfo
  {pages} {062108} (\bibinfo {year} {2018})}\BibitemShut {NoStop}%
\bibitem [{\citenamefont {Reich}\ \emph {et~al.}(2015)\citenamefont {Reich},
  \citenamefont {Katz},\ and\ \citenamefont {Koch}}]{task5}%
  \BibitemOpen
  \bibfield  {author} {\bibinfo {author} {\bibfnamefont {D.~M.}\ \bibnamefont
  {Reich}}, \bibinfo {author} {\bibfnamefont {N.}~\bibnamefont {Katz}}, \ and\
  \bibinfo {author} {\bibfnamefont {C.~P.}\ \bibnamefont {Koch}},\ }\href
  {https://www.nature.com/articles/srep12430} {\bibfield  {journal} {\bibinfo
  {journal} {Scientific Reports}\ }\textbf {\bibinfo {volume} {5}},\ \bibinfo
  {pages} {12430} (\bibinfo {year} {2015})}\BibitemShut {NoStop}%
\bibitem [{\citenamefont {devetak}\ and\ \citenamefont {Winter}(2004)}]{M2}%
  \BibitemOpen
  \bibfield  {author} {\bibinfo {author} {\bibfnamefont {I.}~\bibnamefont
  {devetak}}\ and\ \bibinfo {author} {\bibfnamefont {A.}~\bibnamefont
  {Winter}},\ }\href {http://ieeexplore.ieee.org/document/1362905/} {\bibfield
  {journal} {\bibinfo  {journal} {IEEE Transactions on Information Theory}\
  }\textbf {\bibinfo {volume} {50}},\ \bibinfo {pages} {3183} (\bibinfo {year}
  {2004})}\BibitemShut {NoStop}%
\bibitem [{\citenamefont {Horodecki}\ \emph {et~al.}(2009)\citenamefont
  {Horodecki}, \citenamefont {Horodecki}, \citenamefont {Horodecki},\ and\
  \citenamefont {Horodecki}}]{Re1}%
  \BibitemOpen
  \bibfield  {author} {\bibinfo {author} {\bibfnamefont {R.}~\bibnamefont
  {Horodecki}}, \bibinfo {author} {\bibfnamefont {P.}~\bibnamefont
  {Horodecki}}, \bibinfo {author} {\bibfnamefont {M.}~\bibnamefont
  {Horodecki}}, \ and\ \bibinfo {author} {\bibfnamefont {K.}~\bibnamefont
  {Horodecki}},\ }\href {\doibase 10.1103/RevModPhys.81.865} {\bibfield
  {journal} {\bibinfo  {journal} {Rev. Mod. Phys.}\ }\textbf {\bibinfo {volume}
  {81}},\ \bibinfo {pages} {865} (\bibinfo {year} {2009})}\BibitemShut
  {NoStop}%
\bibitem [{\citenamefont {{Plenio}}\ and\ \citenamefont
  {{Virmani}}(2005)}]{Re2}%
  \BibitemOpen
  \bibfield  {author} {\bibinfo {author} {\bibfnamefont {M.~B.}\ \bibnamefont
  {{Plenio}}}\ and\ \bibinfo {author} {\bibfnamefont {S.}~\bibnamefont
  {{Virmani}}},\ }\href@noop {} {\bibfield  {journal} {\bibinfo  {journal}
  {eprint arXiv:quant-ph/0504163}\ } (\bibinfo {year} {2005})},\ \Eprint
  {http://arxiv.org/abs/quant-ph/0504163} {quant-ph/0504163} \BibitemShut
  {NoStop}%
\bibitem [{\citenamefont {Streltsov}\ \emph {et~al.}(2017)\citenamefont
  {Streltsov}, \citenamefont {Adesso},\ and\ \citenamefont {Plenio}}]{Re4}%
  \BibitemOpen
  \bibfield  {author} {\bibinfo {author} {\bibfnamefont {A.}~\bibnamefont
  {Streltsov}}, \bibinfo {author} {\bibfnamefont {G.}~\bibnamefont {Adesso}}, \
  and\ \bibinfo {author} {\bibfnamefont {M.~B.}\ \bibnamefont {Plenio}},\
  }\href {\doibase 10.1103/RevModPhys.89.041003} {\bibfield  {journal}
  {\bibinfo  {journal} {Rev. Mod. Phys.}\ }\textbf {\bibinfo {volume} {89}},\
  \bibinfo {pages} {041003} (\bibinfo {year} {2017})}\BibitemShut {NoStop}%
\bibitem [{\citenamefont {Barrett}(2007)}]{Re3}%
  \BibitemOpen
  \bibfield  {author} {\bibinfo {author} {\bibfnamefont {J.}~\bibnamefont
  {Barrett}},\ }\href {\doibase 10.1103/PhysRevA.75.032304} {\bibfield
  {journal} {\bibinfo  {journal} {Phys. Rev. A}\ }\textbf {\bibinfo {volume}
  {75}},\ \bibinfo {pages} {032304} (\bibinfo {year} {2007})}\BibitemShut
  {NoStop}%
\bibitem [{\citenamefont {Brunner}\ \emph {et~al.}(2014)\citenamefont
  {Brunner}, \citenamefont {Cavalcanti}, \citenamefont {Pironio}, \citenamefont
  {Scarani},\ and\ \citenamefont {Wehner}}]{Re7}%
  \BibitemOpen
  \bibfield  {author} {\bibinfo {author} {\bibfnamefont {N.}~\bibnamefont
  {Brunner}}, \bibinfo {author} {\bibfnamefont {D.}~\bibnamefont {Cavalcanti}},
  \bibinfo {author} {\bibfnamefont {S.}~\bibnamefont {Pironio}}, \bibinfo
  {author} {\bibfnamefont {V.}~\bibnamefont {Scarani}}, \ and\ \bibinfo
  {author} {\bibfnamefont {S.}~\bibnamefont {Wehner}},\ }\href {\doibase
  10.1103/RevModPhys.86.419} {\bibfield  {journal} {\bibinfo  {journal} {Rev.
  Mod. Phys.}\ }\textbf {\bibinfo {volume} {86}},\ \bibinfo {pages} {419}
  (\bibinfo {year} {2014})}\BibitemShut {NoStop}%
\bibitem [{\citenamefont {Weedbrook}\ \emph {et~al.}(2012)\citenamefont
  {Weedbrook}, \citenamefont {Pirandola}, \citenamefont {Garc\'{\i}a-Patr\'on},
  \citenamefont {Cerf}, \citenamefont {Ralph}, \citenamefont {Shapiro},\ and\
  \citenamefont {Lloyd}}]{Re8}%
  \BibitemOpen
  \bibfield  {author} {\bibinfo {author} {\bibfnamefont {C.}~\bibnamefont
  {Weedbrook}}, \bibinfo {author} {\bibfnamefont {S.}~\bibnamefont
  {Pirandola}}, \bibinfo {author} {\bibfnamefont {R.}~\bibnamefont
  {Garc\'{\i}a-Patr\'on}}, \bibinfo {author} {\bibfnamefont {N.~J.}\
  \bibnamefont {Cerf}}, \bibinfo {author} {\bibfnamefont {T.~C.}\ \bibnamefont
  {Ralph}}, \bibinfo {author} {\bibfnamefont {J.~H.}\ \bibnamefont {Shapiro}},
  \ and\ \bibinfo {author} {\bibfnamefont {S.}~\bibnamefont {Lloyd}},\ }\href
  {\doibase 10.1103/RevModPhys.84.621} {\bibfield  {journal} {\bibinfo
  {journal} {Rev. Mod. Phys.}\ }\textbf {\bibinfo {volume} {84}},\ \bibinfo
  {pages} {621} (\bibinfo {year} {2012})}\BibitemShut {NoStop}%
\bibitem [{\citenamefont {Goold}\ \emph {et~al.}(2016)\citenamefont {Goold},
  \citenamefont {Huber}, \citenamefont {Riera}, \citenamefont {del Rio},\ and\
  \citenamefont {Skrzypczyk}}]{Re5}%
  \BibitemOpen
  \bibfield  {author} {\bibinfo {author} {\bibfnamefont {J.}~\bibnamefont
  {Goold}}, \bibinfo {author} {\bibfnamefont {M.}~\bibnamefont {Huber}},
  \bibinfo {author} {\bibfnamefont {A.}~\bibnamefont {Riera}}, \bibinfo
  {author} {\bibfnamefont {L.}~\bibnamefont {del Rio}}, \ and\ \bibinfo
  {author} {\bibfnamefont {P.}~\bibnamefont {Skrzypczyk}},\ }\href
  {http://stacks.iop.org/1751-8121/49/i=14/a=143001} {\bibfield  {journal}
  {\bibinfo  {journal} {Journal of Physics A: Mathematical and Theoretical}\
  }\textbf {\bibinfo {volume} {49}},\ \bibinfo {pages} {143001} (\bibinfo
  {year} {2016})}\BibitemShut {NoStop}%
\bibitem [{\citenamefont {Gour}\ \emph {et~al.}(2015)\citenamefont {Gour},
  \citenamefont {Müller}, \citenamefont {Narasimhachar}, \citenamefont
  {Spekkens},\ and\ \citenamefont {Halpern}}]{Re6}%
  \BibitemOpen
  \bibfield  {author} {\bibinfo {author} {\bibfnamefont {G.}~\bibnamefont
  {Gour}}, \bibinfo {author} {\bibfnamefont {M.~P.}\ \bibnamefont {Müller}},
  \bibinfo {author} {\bibfnamefont {V.}~\bibnamefont {Narasimhachar}}, \bibinfo
  {author} {\bibfnamefont {R.~W.}\ \bibnamefont {Spekkens}}, \ and\ \bibinfo
  {author} {\bibfnamefont {N.~Y.}\ \bibnamefont {Halpern}},\ }\href {\doibase
  https://doi.org/10.1016/j.physrep.2015.04.003} {\bibfield  {journal}
  {\bibinfo  {journal} {Physics Reports}\ }\textbf {\bibinfo {volume} {583}},\
  \bibinfo {pages} {1 } (\bibinfo {year} {2015})},\ \bibinfo {note} {the
  resource theory of informational nonequilibrium in
  thermodynamics}\BibitemShut {NoStop}%
\bibitem [{\citenamefont {{Wakakuwa}}(2017)}]{rtnm1}%
  \BibitemOpen
  \bibfield  {author} {\bibinfo {author} {\bibfnamefont {E.}~\bibnamefont
  {{Wakakuwa}}},\ }\href@noop {} {\bibfield  {journal} {\bibinfo  {journal}
  {ArXiv e-prints}\ } (\bibinfo {year} {2017})},\ \Eprint
  {http://arxiv.org/abs/1709.07248} {arXiv:1709.07248 [quant-ph]} \BibitemShut
  {NoStop}%
\bibitem [{\citenamefont {Vacchini}\ \emph {et~al.}(2011)\citenamefont
  {Vacchini}, \citenamefont {Smirne}, \citenamefont {Laine}, \citenamefont
  {Piilo},\ and\ \citenamefont {Breuer}}]{example1}%
  \BibitemOpen
  \bibfield  {author} {\bibinfo {author} {\bibfnamefont {B.}~\bibnamefont
  {Vacchini}}, \bibinfo {author} {\bibfnamefont {A.}~\bibnamefont {Smirne}},
  \bibinfo {author} {\bibfnamefont {E.-M.}\ \bibnamefont {Laine}}, \bibinfo
  {author} {\bibfnamefont {J.}~\bibnamefont {Piilo}}, \ and\ \bibinfo {author}
  {\bibfnamefont {H.-P.}\ \bibnamefont {Breuer}},\ }\href
  {http://stacks.iop.org/1367-2630/13/i=9/a=093004} {\bibfield  {journal}
  {\bibinfo  {journal} {New Journal of Physics}\ }\textbf {\bibinfo {volume}
  {13}},\ \bibinfo {pages} {093004} (\bibinfo {year} {2011})}\BibitemShut
  {NoStop}%
\bibitem [{\citenamefont {Brand\~ao}\ and\ \citenamefont {Gour}(2015)}]{rt1}%
  \BibitemOpen
  \bibfield  {author} {\bibinfo {author} {\bibfnamefont {F.~G. S.~L.}\
  \bibnamefont {Brand\~ao}}\ and\ \bibinfo {author} {\bibfnamefont
  {G.}~\bibnamefont {Gour}},\ }\href {\doibase 10.1103/PhysRevLett.115.070503}
  {\bibfield  {journal} {\bibinfo  {journal} {Phys. Rev. Lett.}\ }\textbf
  {\bibinfo {volume} {115}},\ \bibinfo {pages} {070503} (\bibinfo {year}
  {2015})}\BibitemShut {NoStop}%
\bibitem [{\citenamefont {Pollock}\ \emph {et~al.}(2018)\citenamefont
  {Pollock}, \citenamefont {Rodr\'{\i}guez-Rosario}, \citenamefont
  {Frauenheim}, \citenamefont {Paternostro},\ and\ \citenamefont
  {Modi}}]{modi}%
  \BibitemOpen
  \bibfield  {author} {\bibinfo {author} {\bibfnamefont {F.~A.}\ \bibnamefont
  {Pollock}}, \bibinfo {author} {\bibfnamefont {C.}~\bibnamefont
  {Rodr\'{\i}guez-Rosario}}, \bibinfo {author} {\bibfnamefont {T.}~\bibnamefont
  {Frauenheim}}, \bibinfo {author} {\bibfnamefont {M.}~\bibnamefont
  {Paternostro}}, \ and\ \bibinfo {author} {\bibfnamefont {K.}~\bibnamefont
  {Modi}},\ }\href {\doibase 10.1103/PhysRevLett.120.040405} {\bibfield
  {journal} {\bibinfo  {journal} {Phys. Rev. Lett.}\ }\textbf {\bibinfo
  {volume} {120}},\ \bibinfo {pages} {040405} (\bibinfo {year}
  {2018})}\BibitemShut {NoStop}%
\bibitem [{\citenamefont {Choi}(1975)}]{choi}%
  \BibitemOpen
  \bibfield  {author} {\bibinfo {author} {\bibfnamefont {M.-D.}\ \bibnamefont
  {Choi}},\ }\href {\doibase https://doi.org/10.1016/0024-3795(75)90075-0}
  {\bibfield  {journal} {\bibinfo  {journal} {Linear Algebra and its
  Applications}\ }\textbf {\bibinfo {volume} {10}},\ \bibinfo {pages} {285 }
  (\bibinfo {year} {1975})}\BibitemShut {NoStop}%
\bibitem [{\citenamefont {Jamiołkowski}(1972)}]{jamil}%
  \BibitemOpen
  \bibfield  {author} {\bibinfo {author} {\bibfnamefont {A.}~\bibnamefont
  {Jamiołkowski}},\ }\href {\doibase
  https://doi.org/10.1016/0034-4877(72)90011-0} {\bibfield  {journal} {\bibinfo
   {journal} {Reports on Mathematical Physics}\ }\textbf {\bibinfo {volume}
  {3}},\ \bibinfo {pages} {275 } (\bibinfo {year} {1972})}\BibitemShut
  {NoStop}%
\bibitem [{\citenamefont {Ando}(2004)}]{trace1}%
  \BibitemOpen
  \bibfield  {author} {\bibinfo {author} {\bibfnamefont {T.}~\bibnamefont
  {Ando}},\ }\href {\doibase https://doi.org/10.1016/j.laa.2003.06.005}
  {\bibfield  {journal} {\bibinfo  {journal} {Linear Algebra and its
  Applications}\ }\textbf {\bibinfo {volume} {379}},\ \bibinfo {pages} {3 }
  (\bibinfo {year} {2004})},\ \bibinfo {note} {special Issue on the Tenth ILAS
  Conference (Auburn, 2002)}\BibitemShut {NoStop}%
\bibitem [{\citenamefont {Wolf}\ \emph {et~al.}(2008)\citenamefont {Wolf},
  \citenamefont {Eisert}, \citenamefont {Cubitt},\ and\ \citenamefont
  {Cirac}}]{wolf1}%
  \BibitemOpen
  \bibfield  {author} {\bibinfo {author} {\bibfnamefont {M.~M.}\ \bibnamefont
  {Wolf}}, \bibinfo {author} {\bibfnamefont {J.}~\bibnamefont {Eisert}},
  \bibinfo {author} {\bibfnamefont {T.~S.}\ \bibnamefont {Cubitt}}, \ and\
  \bibinfo {author} {\bibfnamefont {J.~I.}\ \bibnamefont {Cirac}},\ }\href
  {\doibase 10.1103/PhysRevLett.101.150402} {\bibfield  {journal} {\bibinfo
  {journal} {Phys. Rev. Lett.}\ }\textbf {\bibinfo {volume} {101}},\ \bibinfo
  {pages} {150402} (\bibinfo {year} {2008})}\BibitemShut {NoStop}%
\bibitem [{\citenamefont {Wolf}\ and\ \citenamefont {Cirac}(2008)}]{wolf2}%
  \BibitemOpen
  \bibfield  {author} {\bibinfo {author} {\bibfnamefont {M.~M.}\ \bibnamefont
  {Wolf}}\ and\ \bibinfo {author} {\bibfnamefont {J.~I.}\ \bibnamefont
  {Cirac}},\ }\href {\doibase 10.1007/s00220-008-0411-y} {\bibfield  {journal}
  {\bibinfo  {journal} {Communications in Mathematical Physics}\ }\textbf
  {\bibinfo {volume} {279}},\ \bibinfo {pages} {147} (\bibinfo {year}
  {2008})}\BibitemShut {NoStop}%
\bibitem [{\citenamefont {Krein}\ and\ \citenamefont {Milman}(1940)}]{M1}%
  \BibitemOpen
  \bibfield  {author} {\bibinfo {author} {\bibfnamefont {M.}~\bibnamefont
  {Krein}}\ and\ \bibinfo {author} {\bibfnamefont {D.}~\bibnamefont {Milman}},\
  }\href {http://eudml.org/doc/219061} {\bibfield  {journal} {\bibinfo
  {journal} {Studia Mathematica}\ }\textbf {\bibinfo {volume} {9}},\ \bibinfo
  {pages} {133} (\bibinfo {year} {1940})}\BibitemShut {NoStop}%
\bibitem [{\citenamefont {Brand{\~a}o}\ and\ \citenamefont
  {Plenio}(2010)}]{rob1}%
  \BibitemOpen
  \bibfield  {author} {\bibinfo {author} {\bibfnamefont {F.~G. S.~L.}\
  \bibnamefont {Brand{\~a}o}}\ and\ \bibinfo {author} {\bibfnamefont {M.~B.}\
  \bibnamefont {Plenio}},\ }\href {\doibase 10.1007/s00220-010-1003-1}
  {\bibfield  {journal} {\bibinfo  {journal} {Communications in Mathematical
  Physics}\ }\textbf {\bibinfo {volume} {295}},\ \bibinfo {pages} {829}
  (\bibinfo {year} {2010})}\BibitemShut {NoStop}%
\bibitem [{\citenamefont {Vidal}\ and\ \citenamefont {Tarrach}(1999)}]{rob2}%
  \BibitemOpen
  \bibfield  {author} {\bibinfo {author} {\bibfnamefont {G.}~\bibnamefont
  {Vidal}}\ and\ \bibinfo {author} {\bibfnamefont {R.}~\bibnamefont
  {Tarrach}},\ }\href {\doibase 10.1103/PhysRevA.59.141} {\bibfield  {journal}
  {\bibinfo  {journal} {Phys. Rev. A}\ }\textbf {\bibinfo {volume} {59}},\
  \bibinfo {pages} {141} (\bibinfo {year} {1999})}\BibitemShut {NoStop}%
\bibitem [{\citenamefont {Harrow}\ and\ \citenamefont {Nielsen}(2003)}]{rob3}%
  \BibitemOpen
  \bibfield  {author} {\bibinfo {author} {\bibfnamefont {A.~W.}\ \bibnamefont
  {Harrow}}\ and\ \bibinfo {author} {\bibfnamefont {M.~A.}\ \bibnamefont
  {Nielsen}},\ }\href {\doibase 10.1103/PhysRevA.68.012308} {\bibfield
  {journal} {\bibinfo  {journal} {Phys. Rev. A}\ }\textbf {\bibinfo {volume}
  {68}},\ \bibinfo {pages} {012308} (\bibinfo {year} {2003})}\BibitemShut
  {NoStop}%
\bibitem [{\citenamefont {Napoli}\ \emph {et~al.}(2016)\citenamefont {Napoli},
  \citenamefont {Bromley}, \citenamefont {Cianciaruso}, \citenamefont {Piani},
  \citenamefont {Johnston},\ and\ \citenamefont {Adesso}}]{rob4}%
  \BibitemOpen
  \bibfield  {author} {\bibinfo {author} {\bibfnamefont {C.}~\bibnamefont
  {Napoli}}, \bibinfo {author} {\bibfnamefont {T.~R.}\ \bibnamefont {Bromley}},
  \bibinfo {author} {\bibfnamefont {M.}~\bibnamefont {Cianciaruso}}, \bibinfo
  {author} {\bibfnamefont {M.}~\bibnamefont {Piani}}, \bibinfo {author}
  {\bibfnamefont {N.}~\bibnamefont {Johnston}}, \ and\ \bibinfo {author}
  {\bibfnamefont {G.}~\bibnamefont {Adesso}},\ }\href {\doibase
  10.1103/PhysRevLett.116.150502} {\bibfield  {journal} {\bibinfo  {journal}
  {Phys. Rev. Lett.}\ }\textbf {\bibinfo {volume} {116}},\ \bibinfo {pages}
  {150502} (\bibinfo {year} {2016})}\BibitemShut {NoStop}%
\bibitem [{\citenamefont {Piani}\ \emph {et~al.}(2016)\citenamefont {Piani},
  \citenamefont {Cianciaruso}, \citenamefont {Bromley}, \citenamefont {Napoli},
  \citenamefont {Johnston},\ and\ \citenamefont {Adesso}}]{rob5}%
  \BibitemOpen
  \bibfield  {author} {\bibinfo {author} {\bibfnamefont {M.}~\bibnamefont
  {Piani}}, \bibinfo {author} {\bibfnamefont {M.}~\bibnamefont {Cianciaruso}},
  \bibinfo {author} {\bibfnamefont {T.~R.}\ \bibnamefont {Bromley}}, \bibinfo
  {author} {\bibfnamefont {C.}~\bibnamefont {Napoli}}, \bibinfo {author}
  {\bibfnamefont {N.}~\bibnamefont {Johnston}}, \ and\ \bibinfo {author}
  {\bibfnamefont {G.}~\bibnamefont {Adesso}},\ }\href {\doibase
  10.1103/PhysRevA.93.042107} {\bibfield  {journal} {\bibinfo  {journal} {Phys.
  Rev. A}\ }\textbf {\bibinfo {volume} {93}},\ \bibinfo {pages} {042107}
  (\bibinfo {year} {2016})}\BibitemShut {NoStop}%
\bibitem [{\citenamefont {Vidal}\ and\ \citenamefont {Werner}(2002)}]{rob6}%
  \BibitemOpen
  \bibfield  {author} {\bibinfo {author} {\bibfnamefont {G.}~\bibnamefont
  {Vidal}}\ and\ \bibinfo {author} {\bibfnamefont {R.~F.}\ \bibnamefont
  {Werner}},\ }\href {\doibase 10.1103/PhysRevA.65.032314} {\bibfield
  {journal} {\bibinfo  {journal} {Phys. Rev. A}\ }\textbf {\bibinfo {volume}
  {65}},\ \bibinfo {pages} {032314} (\bibinfo {year} {2002})}\BibitemShut
  {NoStop}%
\end{thebibliography}%

\end{document}